%% file: conference_101719.tex
\documentclass[conference]{IEEEtran}
\IEEEoverridecommandlockouts
\usepackage{cite}
\usepackage{amsmath,amssymb,amsfonts}
\usepackage{algorithmic}
\usepackage{graphicx}
\usepackage{textcomp}
\usepackage{xcolor}
\usepackage{booktabs}
\usepackage{multirow}
\usepackage{multicol}
\usepackage{stfloats}
\usepackage{hyperref}
\usepackage{float}
\usepackage{siunitx}
\usepackage{url}

\usepackage{breakurl}
\def\BibTeX{{\rm B\kern-.05em{\sc i\kern-.025em b}\kern-.08em
    T\kern-.1667em\lower.7ex\hbox{E}\kern-.125emX}}
\begin{document}

\newcommand{\CR}[1]{\textcolor{black}{#1}}

\title{GauRast: Enhancing GPU Triangle Rasterizers to Accelerate 3D Gaussian Splatting}

\newcommand{\NOTE}[1]{\textcolor{red}{#1}}


\author{\IEEEauthorblockN{Sixu Li$^{1,2}$, Ben Keller$^2$, Yingyan (Celine) Lin$^{1}$, and Brucek Khailany$^3$}
\IEEEauthorblockA{\textit{$^1$School of Computer Science, Georgia Institute of Technology, Atlanta, GA, USA} \\
\textit{
$^2$NVIDIA, Santa Clara, CA, USA}\\
\textit{
$^3$NVIDIA, Austin, TX, USA}\\
\{sli941, celine.lin\}@gatech.edu, \{benk, bkhailany\}@nvidia.com}
}

\maketitle

\input{tex/0-abs}

\input{tex/1-intro}
\input{tex/2-backgroundprof}

\input{tex/3-Motivation}
\input{tex/4-method}
\input{tex/5-evaluation}

\CR{\section{Acknowledgment}
The authors would like to thank Thierry Tambe and Rangharajan Venkatesan for their help with the HLS-based hardware design flow; Ben Eckart, Sifei Liu, and Marco Salvi for their participation in discussions on the 3DGS algorithm; and the anonymous reviewers for their valuable questions and constructive feedback.
This work is supported by an internship at NVIDIA Research, the National Science Foundation (NSF) Computing and Communication Foundations (CCF) program (Award ID: 2312758), and the Department of Health and Human Services Advanced Research Projects Agency for Health (ARPA-H) under Award Number AY1AX000003.}

\Urlmuskip=0mu plus 1mu\relax
\bibliographystyle{IEEEtranS}
\bibliography{refs}

\end{document}

%% file: tex/0-abs.tex
\begin{abstract}

3D intelligence leverages rich 3D features and stands as a promising frontier in AI, with 3D rendering fundamental to many downstream applications. 3D Gaussian Splatting (3DGS), an emerging high-quality 3D rendering method, requires significant computation, making real-time execution on existing GPU-equipped edge devices infeasible.
Previous efforts to accelerate 3DGS rely on dedicated accelerators that require substantial integration overhead and hardware costs. This work proposes an acceleration strategy that leverages the similarities between the 3DGS pipeline and the highly optimized conventional graphics pipeline in modern GPUs. Instead of developing a dedicated accelerator, we enhance existing GPU rasterizer hardware to efficiently support 3DGS operations. Our results demonstrate a 23$\times$ increase in processing speed and a 24$\times$ reduction in energy consumption \CR{for the dominant rasterization operator in the 3DGS pipeline. These improvements yield a 6$\times$ and 4$\times$ faster end-to-end runtime for the original 3DGS algorithm and the latest efficiency-improved pipeline, respectively, achieving rendering speeds of 24 FPS and 46 FPS.} These enhancements incur only a minimal area overhead of 0.2\% relative to the entire SoC chip area, underscoring the practicality and efficiency of our approach for enabling 3DGS rendering on resource-constrained platforms.
\end{abstract}

\begin{IEEEkeywords}
\CR{Architecture \& System Design}, Graphics Processors, 3D Gaussian Splatting, \CR{Neural Rendering}
\end{IEEEkeywords}

%% file: tex/1-intro.tex
\section{Introduction}

\begin{figure}[!t]
\centering
\includegraphics[width=1.0\linewidth]{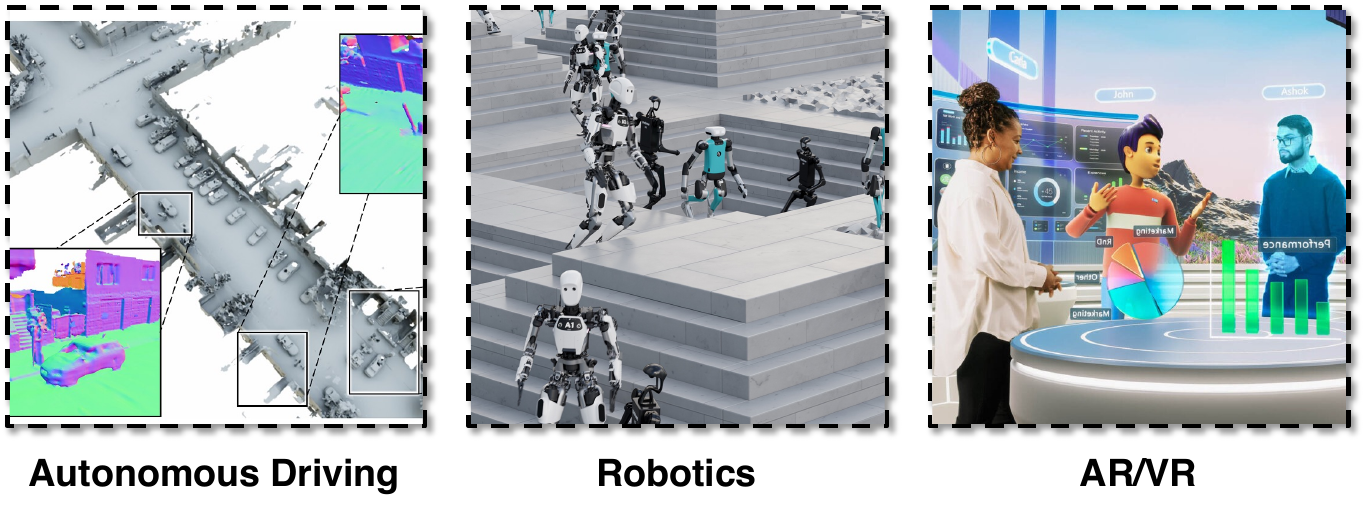}
\vspace{-2.5em}
\caption{Representative examples of 3D intelligent applications, including autonomous driving, robotics, and augmented/virtual reality~\cite{intro-selfdriving, intro-robotics, intro-arvr}.}
\label{fig:intro}
\end{figure}

3D intelligence is set to become the next major frontier in AI by leveraging rich 3D features to enhance understanding and interaction within complex environments. As Prof. Fei-Fei Li, co-founder of ImageNet, emphasized, \textit{``...we need spatially intelligent AI that can model the world and reason about objects, places, and interactions in 3D space and time...''}~\cite{worldlabs}. This underscores the importance of 3D intelligent applications such as autonomous driving~\cite{xie2023s}, robotics~\cite{shen2023F3RM}, and augmented/virtual reality (AR/VR)~\cite{meta_telepresence} shown in Fig.~\ref{fig:intro}. A foundational task within 3D intelligence is 3D scene rendering, which provides essential features for downstream AI tasks that require spatial information.

Recently, 3D Gaussian Splatting (3DGS)~\cite{kerbl20233d} has emerged as the leading method for 3D scene rendering. As summarized in Table~\ref{tab:intro}, unlike previous techniques like triangle meshes~\cite{shirley2009fundamentals} and neural radiance fields (NeRF)~\cite{mildenhall2021nerf}, 3DGS maps 3D scenes onto a set of Gaussian balls, offering superior rendering quality and automated scene reconstruction. Thanks to these advantages, 3DGS has been rapidly adopted in various 3D applications, including autonomous driving~\cite{zhou2024drivinggaussian, huang2024s3gaussian, qi2024gspr}, robotics~\cite{hu2025cg, matsuki2024gaussian, abouphysically}, and AR/VR~\cite{qian20243dgs, ma20243d, yuan2024gavatar}.

\begin{table}[!t]
\vspace{-1em}
    \caption{Comparison of Rendering Methodologies}
    \vspace{-0.7em}
  \centering
    \resizebox{1\linewidth}{!}
    {
      \begin{tabular}{c|ccc}
      
      \toprule[0.1pt]
       & \textbf{Triangle Mesh~\cite{shirley2009fundamentals}} & \textbf{NeRF~\cite{mildenhall2021nerf}} & \textbf{3D Gaussian~\cite{kerbl20233d}} \\
      \midrule
      Scene Reconstruction & Manual & Automatic & Automatic \\
      \midrule
      Rendering Quality & Manually Decided & High & Very High \\
      \midrule
      Rendering Speed on GPU~\cite{orinnx} & Fast & Slow & Medium  \\
      \bottomrule[0.1pt]
      \end{tabular}
      }
      \vspace{-1.7em}
    \label{tab:intro}
\end{table}

Despite its robust performance on high-powered ($\geq200$~W) GPUs~\cite{A6000GPU}, achieving real-time rendering of $\geq30$ frames per second (FPS), 3D Gaussian Splatting cannot achieve high framerates on edge computing platforms with GPUs, such as those with power limitations ($\leq10$ W) like the NVIDIA Jetson Orin NX~\cite{orinnx}. These platforms are increasingly crucial due to the growing demand for 3D processing in mobile and embedded systems~\cite{intro-selfdriving, intro-arvr}. Specifically, 3DGS achieves only 2-5 FPS on these platforms~\cite{orinnx} with commonly used real-world, large-scale datasets~\cite{barron2022mip}, falling short of the performance requirement for most practical applications. This performance gap poses challenges for deploying advanced 3D intelligence in resource-constrained environments, highlighting the need for hardware acceleration of 3DGS.

Existing efforts to accelerate 3D Gaussian Splatting~\cite{lee2024gscore} and similar rendering pipelines~\cite{li2023instant, feng2024cicero} typically focus on developing dedicated hardware accelerators. While these specialized units can provide performance benefits, they require exclusive hardware resources and often lead to increased system complexity and cost. Moreover, the integration of dedicated accelerators may not be feasible for all edge platforms, especially those with stringent area and power constraints. However, \CR{it is noteworthy that} modern GPUs are already equipped with optimized fixed-function graphic acceleration hardware for triangle meshes. As these fixed-function units also target graphics rendering, their inherent capabilities present an opportunity to leverage existing resources to accelerate 3DGS rendering without introducing significant additional overhead.

\begin{figure}[!t]
\centering
\includegraphics[width=1.0\linewidth]{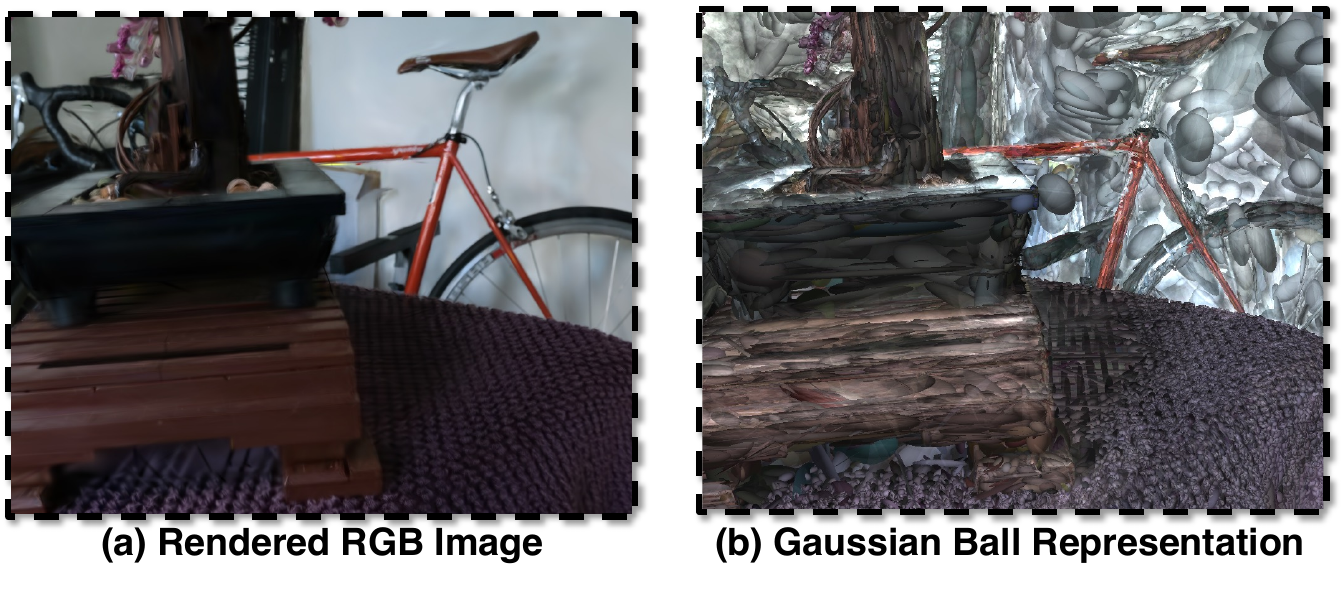}
\vspace{-2em}
\caption{Visualization of a 3D Gaussian representation: (a) Rendered RGB image depicting the scene with realistic color and detail; and (b) the corresponding Gaussian ball representation of the same scene, showing the underlying 3D structure before rendering. Both images are rendered using the 'Bonsai' scene from the NeRF-360~\cite{barron2022mip} dataset.}
\label{fig:pre-vis}
\vspace{-1em}
\end{figure}

To mitigate the challenges associated with designing dedicated memory systems and software stacks for specialized accelerators~\cite{chen2018tvm}, we propose an alternative solution: enhancing the existing graphics units—specifically, the rasterizer for triangle meshes—within GPUs to accelerate 3DGS rendering. By leveraging the capabilities of \CR{those existing fixed function units in} current GPU hardware, we aim to enhance performance without the need for additional dedicated accelerators. We begin with an in-depth profiling of the 3DGS rendering pipeline to identify time-consuming operations and analyze its similarities with operators in the triangle mesh rendering pipeline, which are already supported by the GPU’s native graphic hardware units. This analysis reveals opportunities to adapt and enhance these units to efficiently process 3D Gaussian operations.

\begin{figure*}[b]
\vspace{-1.5em}
\centering
\includegraphics[width=1\linewidth]{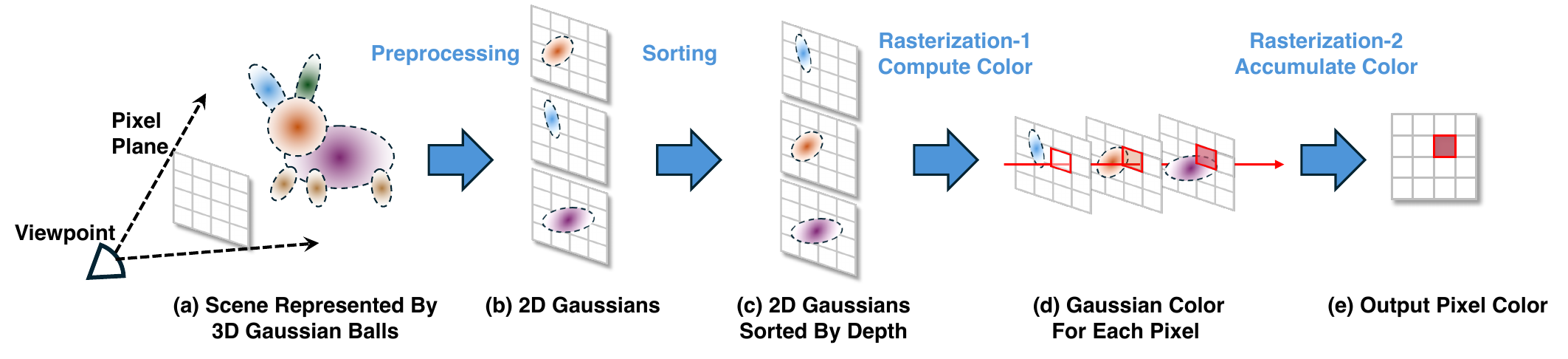}
\vspace{-2.5em}
\caption{Overview of the 3DGS pipeline~\cite{kerbl20233d}: (a) \textit{Scene representation}: The scene is depicted as 3D Gaussian balls, viewed from a specific viewpoint and projected onto a 2D pixel plane. (b) \textit{Preprocessing}: These 3D Gaussians are projected onto the 2D plane, resulting in 2D Gaussian representations. (c) \textit{Sorting}: 2D Gaussians are ordered by depth to ensure the correct rendering sequence and handle occlusion properly. (d) \textit{Initial rasterization}: Colors for each pixel are calculated based on the Gaussians covering that pixel. (e) \textit{Color accumulation}: Colors are accumulated to produce the final pixel color output.}

\label{fig:pre-pip}
\end{figure*}

Building on these insights, we propose and develop an enhanced rasterizer, GauRast, for GPUs that enables the efficient execution of the dominant operator in the 3DGS rendering pipeline while preserving the original capabilities for standard triangle mesh rendering. This design ensures compatibility with existing GPU architectures and minimizes disruptions to conventional workflows. By enhancing existing hardware rather than introducing entirely new components, our approach strikes a balance between performance improvements and resource \CR{overhead}. Our contributions are as follows:

\begin{itemize}

\item We advocate an alternative direction for accelerating 3DGS, the leading algorithm in neural rendering, by enhancing existing GPU units originally dedicated to traditional workloads.

\item We detail comprehensive profiling and analysis of the 3DGS rendering pipeline~\cite{kerbl20233d} to identify the critical operators requiring acceleration, and demonstrate that the bottleneck operator in the 3DGS rendering pipeline shares similarities with the triangle rendering pipeline, which is already well-accelerated by existing GPU's triangle rasterizer.

\item We propose an enhanced rasterizer, named GauRast, to support the dominant operator in the 3DGS rendering pipeline while maintaining its functionality for standard triangle mesh rendering tasks.

\item We present a GauRast hardware prototype that achieves a 23$\times$ speedup and a 24$\times$ improvement in energy efficiency on the target SoC for the dominant operator with the original 3DGS rendering pipeline. This leads to a 6$\times$ end-to-end speedup in the original 3DGS rendering pipeline~\cite{kerbl20233d} and a 4$\times$ end-to-end speedup in the latest efficiency-optimized 3DGS rendering pipeline~\cite{fang2024mini}, achieving 24 FPS and 46 FPS, respectively. Notably, GauRast incurs an area overhead of only 21\% for the enhanced graphics units, corresponding to 0.2\% of the total SoC area.

\end{itemize}

Taken together, our proposals demonstrate a promising approach to enabling real-time 3DGS rendering on future edge devices.

%% file: tex/2-backgroundprof.tex
\section{3DGS and Its Rendering Bottleneck}

\subsection{3D Gaussian Splatting Algorithm}

\textbf{3D Gaussian Scene Representation.} The state-of-the-art (SOTA) method for 3D scene rendering employs a 3D Gaussian-based representation~\cite{kerbl20233d} as illustrated in Fig.~\ref{fig:pre-vis}. This approach strikes a balance between high rendering quality and efficient rendering speed, especially on standard desktop-level GPUs. In this representation, objects are modeled as collections of elliptical 3D Gaussian balls. Each Gaussian ball is characterized by a 3D Gaussian probability density function and an associated color vector. The rendering process distributes colors over the regions covered by the Gaussians' probability densities, collectively forming the complete scene. Rendering an image from a specific viewpoint using this representation (described in Fig.~\ref{fig:pre-pip}) involves three main steps:

\begin{figure}[!t]
\centering
\includegraphics[width=1.0\linewidth]{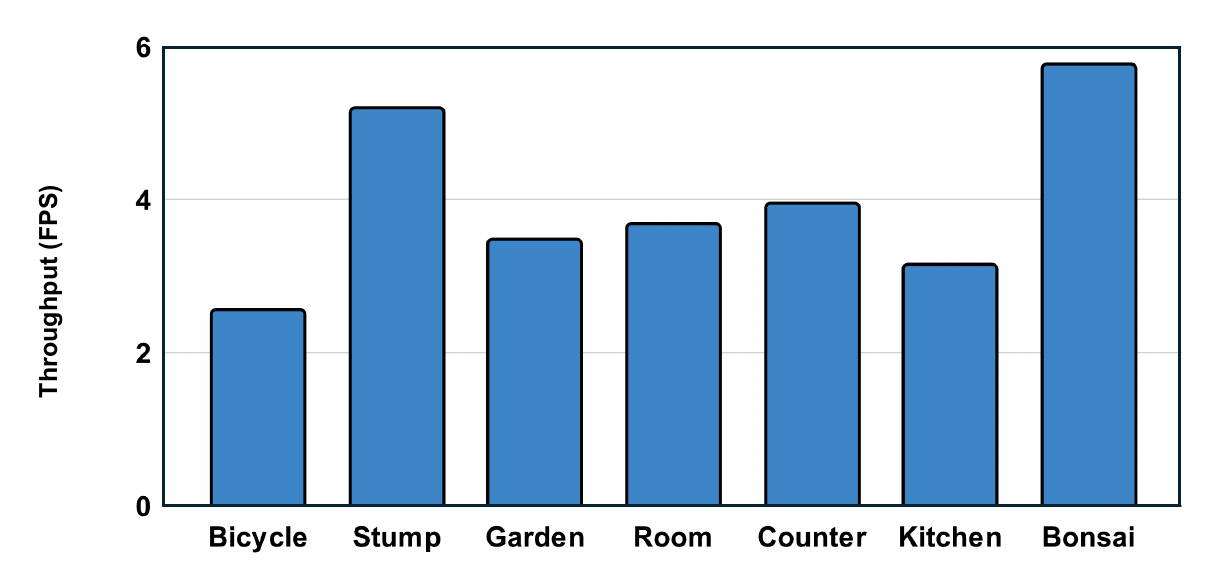}
\vspace{-2.5em}
\caption{Throughput achieved by the 3DGS rendering pipeline~\cite{kerbl20233d} across all seven scenes from the large-scale, real-world NeRF-360 dataset~\cite{barron2022mip}, as measured on the NVIDIA Jetson Orin NX~\cite{orinnx} with a 10W power limit.}
\label{fig:pre-runtime}
\vspace{-1em}
\end{figure}

\textbf{Step 1: Preprocessing.} The first stage of 3DGS rendering, depicted in Fig.~\ref{fig:pre-pip}(a), is preprocessing, which serves three purposes: projecting all 3D Gaussians onto a 2D plane according to the specified viewing position and angle, converting the color vector of each Gaussian to an RGB representation based on the viewing parameters, and computing the depth of each 3D Gaussian relative to the given viewpoint. This step transforms the 3D representation into a set of 2D Gaussians, each characterized by an opacity function \( o \), an assigned RGB color \( c \), a depth value \( d \), a covariance matrix \( \Sigma \) representing the 2D Gaussian probability distribution, and a center point \( \mu \).

\textbf{Step 2: Sorting.} The second stage, as illustrated in Fig.~\ref{fig:pre-pip}(b), involves sorting the projected 2D Gaussians. Due to the potential overlap of Gaussians, each pixel may be influenced by multiple Gaussians. The rendering order impacts their visibility since Gaussians rendered earlier can obscure those rendered later. To maintain correct occlusion relationships and ensure visual consistency, the 2D Gaussians are sorted by their depth values. This depth-based sorting ensures that Gaussians nearer to the viewing position are rendered first, preserving proper occlusion handling.
\textbf{Step 3: Gaussian Rasterization.} The final step, Gaussian rasterization, applies the color of each Gaussian to the pixels it covers, based on opacity and Gaussian probability, and follows the depth order determined in Step 2. As shown in Fig.~\ref{fig:pre-pip}(c), for each Gaussian applied to a specific pixel, the density \( \alpha_{P,i} \) is computed using the formula:
\[
\alpha_{P,i} = o_i e^{-\frac{1}{2}(P-{\mu_i})^T{\Sigma_i}^{-1}(P-{\mu_i})}
\]
where \( P \) represents the pixel coordinate, \( i \) is the index of the Gaussian, and \( o_i \) is the opacity of the Gaussian. This density reflects the contribution of the Gaussian to the pixel. Once the density \( \alpha_{P,i} \) is calculated, the RGB color contribution for that pixel is accumulated as follows, as shown in Fig.~\ref{fig:pre-pip}(d):
\[
\mathbb{C}_P = \sum_{i=1}^{n} T_{P,i} \alpha_{P,i} c_i
\]
The term \( T_{P,i} = \prod_{j=1}^{i-1}(1 - \alpha_{P,j}) \) represents the accumulated density of all previously applied Gaussians at the pixel, accounting for the occlusion effects from preceding Gaussians. This ensures that only visible portions of overlapping Gaussians contribute to the final rendered color in Fig.~\ref{fig:pre-pip}(e).

\subsection{Locating the Bottleneck of 3DGS Rendering}
\label{sec:profiling}

\begin{figure}[!t]
\centering
\includegraphics[width=1.0\linewidth]{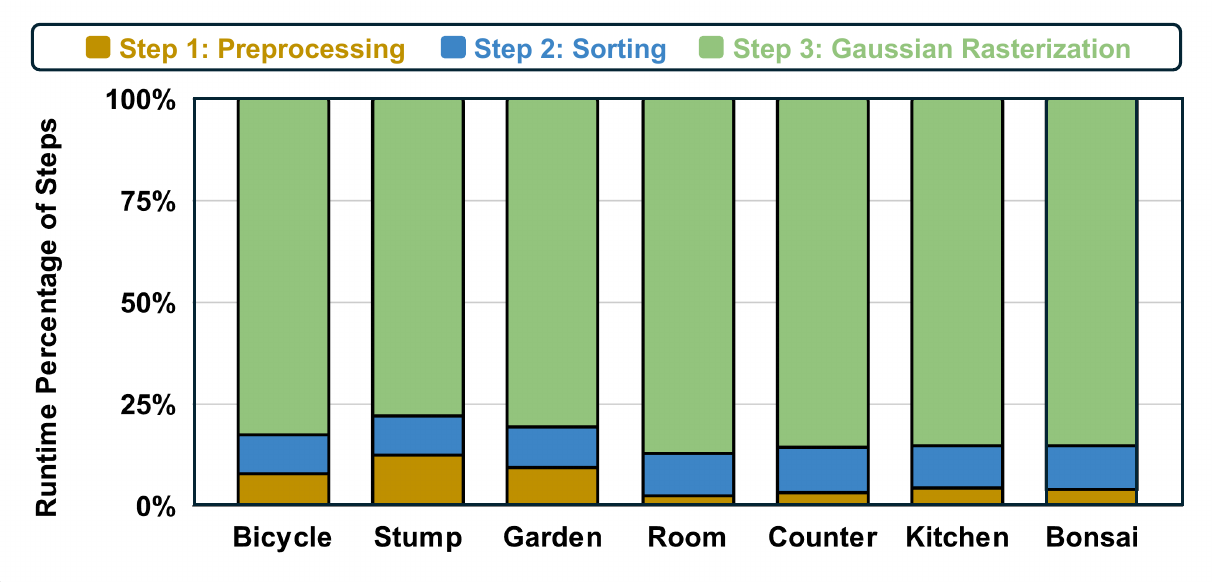}
\vspace{-2.5em}
\caption{Runtime breakdown of the 3DGS rendering pipeline~\cite{kerbl20233d} across all seven scenes from the large-scale, real-world NeRF-360 dataset~\cite{barron2022mip} as measured on the NVIDIA Jetson Orin NX~\cite{orinnx} with a power limit of 10W.}
\label{fig:pre-breakdown}
\vspace{-1em}
\end{figure}

To identify bottlenecks in rendering 3D Gaussians on edge SoCs, we conducted detailed profiling to analyze the runtime distribution across various rendering steps. Profiling was performed on the widely used 3D rendering dataset {NeRF-360}~\cite{barron2022mip}, which includes seven real-world multi-object scenes. Kernel runtimes for the different rendering steps were measured using NVIDIA Nsight Systems~\cite{nsys} on the NVIDIA Jetson Orin NX~\cite{orinnx} operating under a 10W power limit typical for edge devices~\cite{9372829, 10.1145/3604930.3605712}.
The average runtime for rendering all viewpoints in each scene is summarized in Fig.~\ref{fig:pre-runtime}. The results indicate that the edge SoC achieves only 2-5 FPS, necessitating significant speedups before it can be used in real-world applications. The detailed runtime breakdown for each rendering step is shown in Fig.~\ref{fig:pre-breakdown}. Step 3, the Gaussian rasterization stage, dominates the overall runtime, accounting for over 80\% of the total rendering time across all scenes, thereby representing the primary performance bottleneck.

%% file: tex/3-Motivation.tex
\section{Leveraging the GPU Rasterizer for 3DGS}
\subsection{Opportunities in GPU's Rasterizer}
The computational demands of dominant Gaussian rasterization in the 3DGS rendering pipeline have motivated various acceleration efforts~\cite{lee2024gscore}. Current research primarily emphasizes the development of dedicated hardware accelerators for emerging 3D rendering pipelines such as 3DGS~\cite{lee2024gscore} and NeRF~\cite{li2023instant, feng2024cicero}. Although these accelerators offer performance improvements, they can introduce significant trade-offs, including increased complexity in memory systems and software stacks~\cite{chen2018tvm} and the need for exclusive hardware resources. Unlike these dedicated approaches, our strategy focuses on enhancing existing GPU hardware to leverage the built-in capabilities of modern GPUs. This approach enhances 3DGS performance while preserving compatibility with conventional GPU operations and the original programming interface, minimizing disruptions to established workflows and fully utilizing the GPU’s existing data path.
Modern GPUs have dedicated units optimized for triangle mesh rasterization \cite{lifetriangle}. This process involves iterating over primitives (here referring to triangles) and mapping their parameters to display pixels. The pipeline is highly parallelized, relying on fixed-function accelerators for rapid computation \cite{ackermann1996single, chen2002reduce, wang2012more}. Our approach begins with an investigation of the similarities and differences between Gaussian rasterization and triangle rasterization, the latter of which is efficiently supported by these optimized GPU hardware units.

\begin{figure}[!t]
\centering
\includegraphics[width=1.0\linewidth]{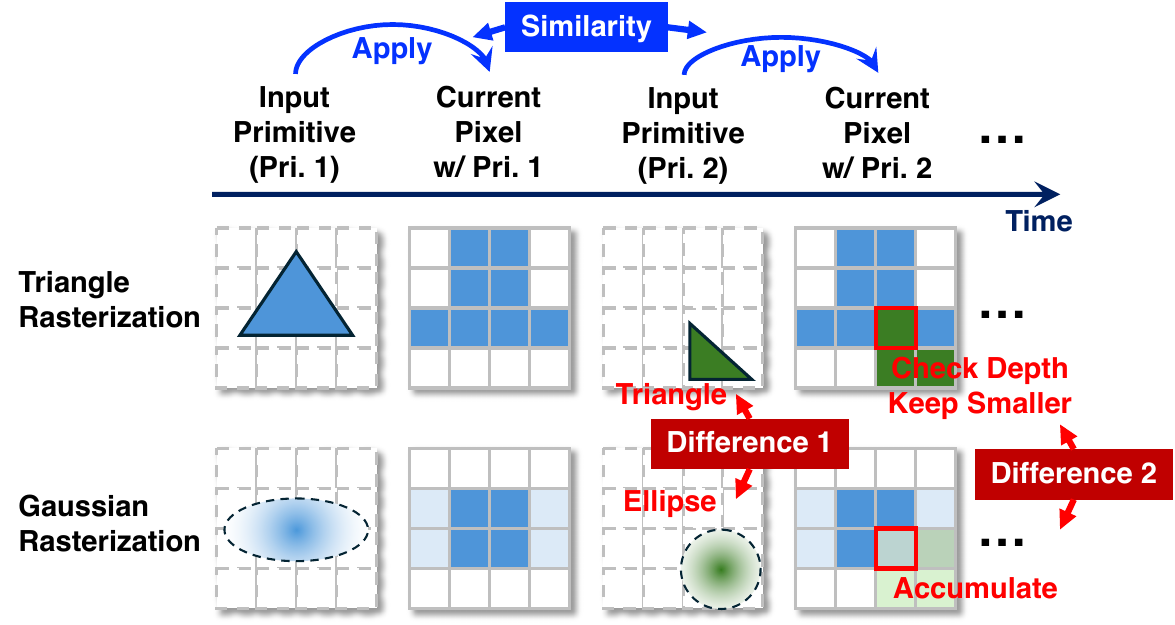}
\vspace{-1.0em}
\caption{Comparison of triangle and Gaussian rasterization. Both techniques sequentially apply primitives to pixels with key differences in primitive type (Difference 1) and reduction algorithm (Difference 2).}
\label{fig:mot-comp}
 \vspace{-1em}
\end{figure}

\subsection{Contrasting Gaussian and Triangle Rasterization Methods}
 \vspace{-0.1em}
3DGS rasterization shares key foundational elements with traditional triangle mesh rasterization. As illustrated in Fig.~\ref{fig:mot-comp}, both processes iterate over primitives—triangles in conventional rendering and Gaussians in 3DGS—and map their respective parameters onto pixels. This similarity in the basic dataflow and computation pattern provides an opportunity to adapt existing GPU rasterizers to support 3D Gaussian operations.

However, there are also differences between the two rasterization processes. First, in triangle rasterization, the algorithm involves determining if a pixel lies within a triangular boundary. In contrast, 3DGS requires calculating Gaussian probability distributions to assess the coverage. This distinction necessitates modifications to the detection function to accommodate elliptical shapes \CR{instead of being limited to triangles}. 
\CR{Second, in triangle mesh rasterization, the reduction method relies on a simple minimum-depth comparison to determine the primitive with the minimum depth that should be applied to each pixel. In contrast, 3DGS aggregates color contributions from multiple overlapping Gaussians. This aggregation increases the complexity of the reduction function, as we need to compute densities and occlusion effects for all the overlapping Gaussians and then sum them together.}
Effectively addressing these differences is essential to adapting current rasterization hardware for 3DGS, ensuring compatibility and preserving performance.

\begin{table}[!t]
    \caption{Computational Primitives for Rasterization}
      \vspace{-0.7em}
  \centering
    \resizebox{1\linewidth}{!}
    {
      \begin{tabular}{c|cc}
      
      \toprule[1pt]
       Subtask & \multirow{2}{*}{\textbf{Triangle Rasterization~\cite{shirley2009fundamentals}}} & \multirow{2}{*}{\textbf{Gaussian Rasterization~\cite{kerbl20233d}}} \\
       (Operator) &&\\

      \midrule
      \multirow{2}{*}{Input} & Vertices' Coordinates & $\Sigma$, $o$, $\mu$, $c$ \\
      &(9 FP Numbers)&(9 FP Numbers)\\
      \midrule
      \multirow{2}{*}{1} & Coordinate Shift & Coordinate Shift \\
      &(ADD, MUL)& (ADD, MUL)\\
      \midrule
      \multirow{2}{*}{2} & Intersection Detection & Gaussian Probability Computation \\
      &(ADD, MUL, DIV)&(ADD, MUL, EXP)\\
      \midrule
      \multirow{2}{*}{3} & UV Weight Computation & Color Weight Computation \\
      &(ADD, MUL)&(ADD, MUL)\\
      \midrule
      \multirow{2}{*}{4} & Min-Depth Color Hold & Color Accumulation \\
      &(ADD, MUL)&(ADD, MUL) \\
      \midrule
      \multirow{2}{*}{Output} & UV Weight, Depth & Accumulated Color \\
      &(3 FP Numbers)&(3 FP Numbers)\\
      \bottomrule[1pt]
      \end{tabular}
      }
      \vspace{-1em}
    \label{tab:meshgaussian}
\end{table}

\subsection{Enhancing the Triangle Rasterizer for 3DGS}
\label{sec:implementationplan}

Building on our identified similarities and differences between Gaussian and triangle rasterization, we propose enhancing the existing triangle rasterizer to support 3DGS. A comparison of the resource requirements for triangle and Gaussian rasterization is summarized in Table~\ref{tab:meshgaussian}. 
As shown in the table, both rasterization processes have identical input and output parameter sizes and follow a similar procedure: initializing pixel storage and then applying primitives to each pixel. Owing to this shared I/O width and access pattern, the existing memory interface can be directly reused from the existing triangle rasterizer.

Regarding computational resources, as highlighted in Table~\ref{tab:meshgaussian}, both processes primarily require multipliers and adders for their core tasks. This similarity allows us to introduce a reconfigurable datapath capable of supporting both triangle and Gaussian primitives with the same hardware resources. However, each primitive type has specific resource requirements: triangle rasterization requires a divider, while Gaussian rasterization necessitates an exponentiation unit. To address these specialized needs, we propose adding dedicated hardware units for these distinct operations, enabling seamless support for both types of rasterization.

%% file: tex/4-method.tex
\section{Proposed GauRast Hardware}

\begin{figure*}[t]
\centering
\includegraphics[width=1\linewidth]{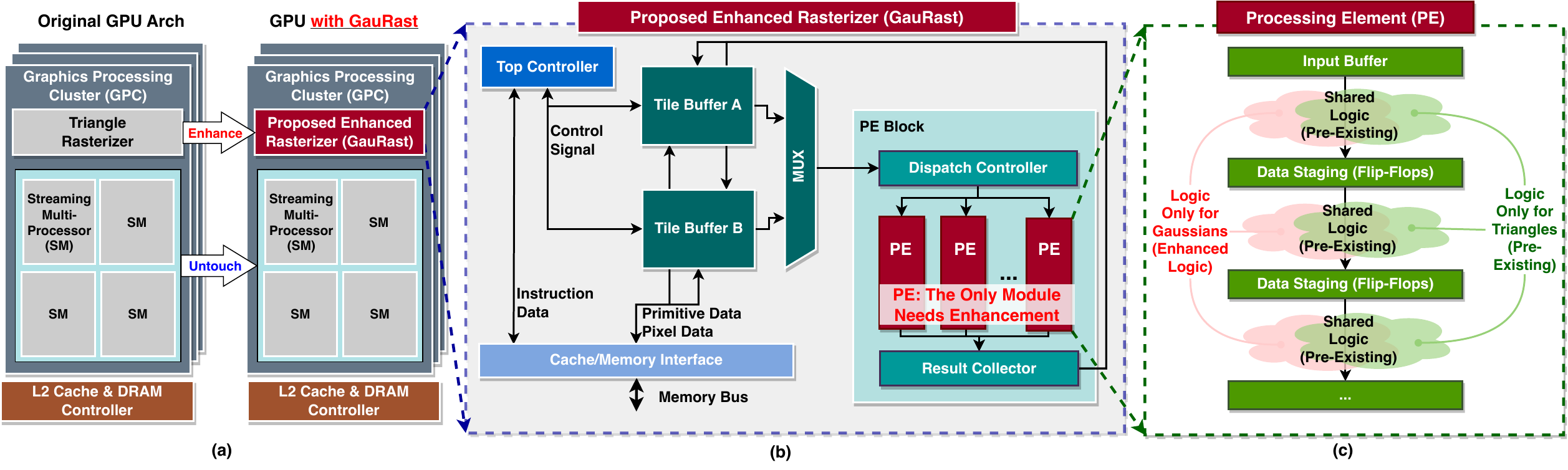}
\vspace{-1em}
\caption{Overview of the proposed hardware architecture: (a) Block diagram of the GPU's Graphics Processing Cluster (GPC) with proposed enhancements. (b) Detailed view of the enhanced rasterizer. (c) Internal structure of each PE, displaying dedicated and shared logic paths for Gaussian and triangle primitives.}

\label{fig:archoverview}
\vspace{-1em}
\end{figure*}

\subsection{Overview of the GPU architecture with GauRast}
Building on the analysis in Section~\ref{sec:implementationplan}, we present the hardware design of GauRast, an enhanced rasterizer that extends existing hardware capabilities to efficiently support 3DGS. Unlike previous approaches that rely on dedicated accelerators, GauRast utilizes the built-in GPU rasterization hardware, specifically the triangle rasterizer, to execute 3DGS rendering tasks. This approach ensures seamless integration with existing GPU workflows, maintaining compatibility with conventional rendering while introducing minimal overhead.

As shown on the left side of Fig.~\ref{fig:archoverview}(a), the GauRast hardware builds upon the Graphics Processing Cluster (GPC) of modern GPU architectures~\cite{voltawp, amperewp}. Each GPC contains multiple Streaming Multiprocessors (SMs) for general-purpose computations and specialized fixed-function units for graphics processing. Within this architecture, we implemented an enhanced rasterizer capable of handling both triangle and Gaussian rasterization, extending the functionality of the original rasterizer that only supports triangle rasterization. As shown in Fig.~\ref{fig:archoverview}(a), our modifications are confined to the rasterizer within the GPC, leaving the SMs untouched and preserving the overall GPU structure. The enhanced rasterizer is optimized to process Gaussian splatting, incorporating new logic paths alongside the existing hardware inside of each Processing Element (PE) as shown in Fig.~\ref{fig:archoverview}(c), thus enabling seamless switching between traditional triangle rendering and Gaussian rasterization.

\vspace{-0.2em}
\subsection{Proposed GauRast Hardware}

The enhanced rasterizer consists of several key components, as shown in Fig.~\ref{fig:archoverview}(b):

\textbf{Tile Buffers (A and B).} These ping-pong buffers store primitives (either Gaussians or triangles) and pixel data, enabling efficient data management during the rendering process. By alternating between Tile Buffer A and Tile Buffer B, GauRast minimizes memory latency and allows concurrent data access for both Gaussian and triangle primitives.

\textbf{PE Block.} The core computation of the rasterization process is handled by the PE Block, which contains multiple PEs. Each PE is tasked with computations required for either Gaussian or triangle primitives. The PE Block operates with a high degree of parallelism, leveraging the intrinsic parallel structure of the rendering task for acceleration.

\textbf{Processing Element (PE).} Each PE supports both Gaussian and triangle rasterization and is equipped with a combination of shared and dedicated logic units, as shown in Fig.~\ref{fig:archoverview}(c). The shared logic handles common operations, while dedicated logic paths manage primitive-specific computations, such as evaluating Gaussian probability distributions for 3DGS rendering and performing depth division for triangle meshes. Importantly, the existing triangle rasterizer includes both shared and triangle-specific logic; thus, only Gaussian-specific logic is added. Compared to a standard triangle rasterizer, our design requires minimal additional hardware in each PE: two adders, one multiplier, and one exponentiation unit for Gaussian support, while reusing nine adders and nine multipliers from the triangle rasterizer. \CR{ Multiplexers were employed to select the input for the hardware units based on the running mode. To save power, input gating is adopted when the modes do not match.} The PE is the only module requiring additional logic to facilitate 3DGS rendering.

The GauRast hardware optimizes performance and resource utilization by reusing existing memory interfaces, such as Tile Buffers and Controllers, and incorporates minimal additional logic for 3DGS tasks. This design ensures that the enhanced rasterizer remains compatible with the existing triangle rendering pipeline, thus minimizing hardware complexity.

\begin{figure}[!t]
\centering
\includegraphics[width=1.0\linewidth]{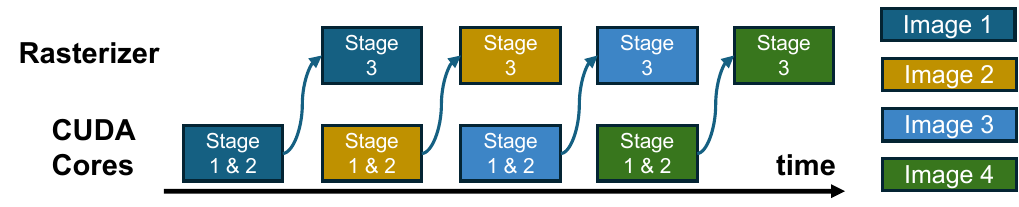}
\vspace{-1.5em}
\caption{\CR{Illustration of CUDA-Collaborative scheduling, where CUDA cores and the rasterizer operate in a pipelined manner to maximize hardware utilization during rendering.}}
\label{fig:coscheduling}
\vspace{-1em}
\end{figure}

\subsection{CUDA-Collaborative Scheduling}
In triangle mesh rasterization, CUDA cores and the hardware rasterizer collaborate to complete the rendering process. Specifically, non-dominant tasks such as 3D-to-2D projection and color querying are handled by the CUDA cores, while the rasterization is managed by the hardware rasterizer~\cite{lifetriangle}. Similarly, we adopt a hybrid scheduling strategy to enhance efficiency, assigning non-dominant operations like preprocessing and sorting to the CUDA cores, while the enhanced rasterizer handles the dominant rasterization workload which is the primary bottleneck in 3DGS rendering. \CR{Specifically, in the 3DGS's 3-stage pipeline, the CUDA-cores process Stages 1-2 of one image before passing it to GauRast for Stage 3, after which the CUDA-cores can begin working on Stages 1-2 of the next image without waiting, as shown in Fig.~\ref{fig:coscheduling}.}

%% file: tex/5-evaluation.tex
\section{Evaluation}

\subsection{Evaluation Setup}
\label{sec:evaluationsetup}

\textbf{Dataset \& Baselines.} To evaluate the processing speedup and efficiency improvement achieved by our proposed GauRast, we conducted experiments using the widely adopted NeRF-360 dataset~\cite{barron2022mip}, a large-scale, real-world dataset using both the original 3DGS algorithm~\cite{kerbl20233d} and its latest efficiency-optimized variant~\cite{fang2024mini}. For baselines, we consider both the NVIDIA Jetson Orin NX SoC~\cite{orinnx}, a representative edge GPU, and the only previously published accelerator proposal for 3DGS, GScore~\cite{lee2024gscore}.

\begin{figure}[!t]
\centering
\includegraphics[width=1.0\linewidth]{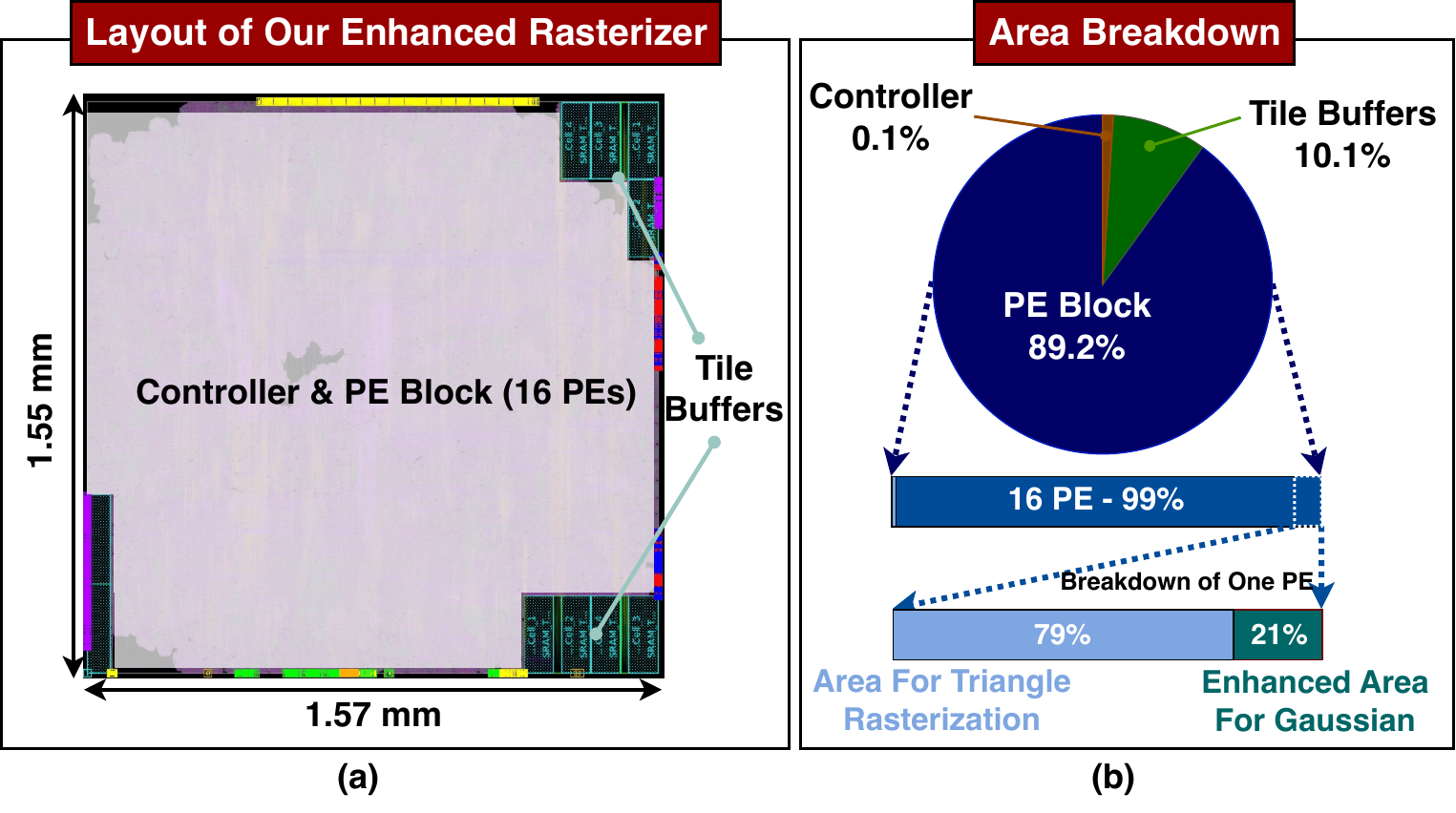}
\vspace{-1.5em}
\caption{Layout and area breakdown of the proposed enhanced rasterizer.}
\label{fig:layout}
\vspace{-1em}
\end{figure}

\textbf{Hardware Implementation.}
We implemented a prototype of the enhanced rasterizer with 16 PEs using C++ and synthesized the design into RTL using Siemens Catapult~\cite{catapult}. This RTL was then synthesized, placed, and routed using Synopsys Fusion Compiler~\cite{fcshell}, targeting a \qty{28}{\nano\meter} CMOS technology node (typical corner, \qty{0.9}{\volt}, \qty{1}{\giga\hertz} clock), following the methodology described in~\cite{khailany2018modular}. \CR{The implementation uses FP32 precision for all computations and leverages Siemens' floating-point IPs~\cite{catapult} to maintain result consistency with the software implementation.} Fig.~\ref{fig:layout} shows the layout and area breakdown of the prototype. The typical power consumption of the prototype is \qty{1.7}{\watt} as reported by Synopsys PrimePower~\cite{primepower} based on post-layout netlists and RTL simulation activities using randomly sampled test images from the target dataset.
To ensure correct implementation, we validated the functional accuracy of both triangle and Gaussian rasterization against the software implementations, confirming that the RTL implementation's rendering output for both triangle rasterization~\cite{tinyrenderer} and 3DGS rendering~\cite{kerbl20233d} matches perfectly without any loss in rendering quality.

\textbf{Simulator Setup.} 
To match the effective area of the triangle rasterizer units in the baseline SoC~\cite{orinnx}, we scaled up our GauRast design to include 15 instances of the 16-PE rasterizer module, totaling 300 PEs.  Consequently, the enhanced portion of the hardware occupies approximately 0.2\% of the total area of the baseline Jetson Orin NX SoC~\cite{orinnx}. We further developed a cycle-accurate simulator for fast evaluation of this scaled-up design. The simulator's runtime and power outputs were validated against the aforementioned RTL simulation results to ensure accuracy and reliability.

\begin{table}[!t]
    \caption{\CR{Absolute Rasterization Runtime w/ and w/o GauRast Compared to CUDA implementations on the NVIDIA Jetson Orin NX~\cite{orinnx}. Evaluations were conducted using both the original 3DGS algorithm~\cite{kerbl20233d}, across all seven scenes of the large-scale, real-world NeRF-360 dataset~\cite{barron2022mip}.}}
      \vspace{-0.7em}
  \centering
    \resizebox{1\linewidth}{!}
    {
      \begin{tabular}{c|ccccccc}
      
      \toprule[1pt]
       Scene & Bicycle & Stump & Garden & Room & Counter & Kitchen & Bonsai\\
      \midrule
        Baseline~\cite{orinnx} & 321 ms &149 ms& 232 ms& 236 ms& 216 ms& 269 ms& 147 ms \\
        \midrule
        GauRast & 15 ms& 6.0 ms& 9.6 ms& 10.5 ms& 9.8 ms& 12.2 ms& 5.5 ms \\
      \bottomrule[1pt]
      \end{tabular}
      }
      \vspace{-1em}
    \label{tab:absoluteruntime}
\end{table}

\begin{figure}[!t]
\centering
\includegraphics[width=1\linewidth]{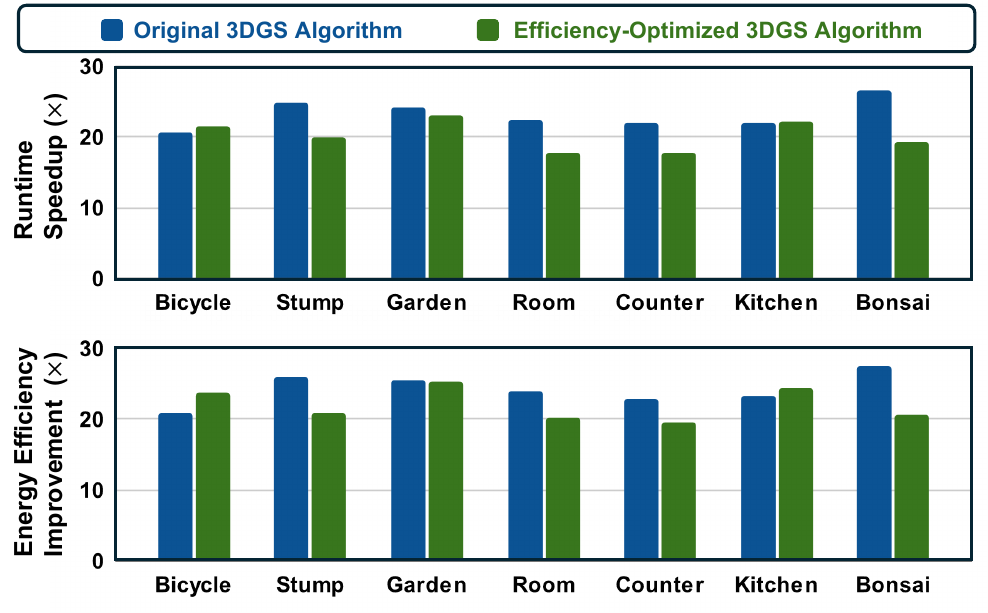}
\vspace{-2em}
\caption{Speedup and energy efficiency improvements achieved by GauRast in Gaussian rasterization, compared to CUDA implementations on the NVIDIA Jetson Orin NX~\cite{orinnx}. Evaluations were conducted using both the original 3DGS algorithm~\cite{kerbl20233d} and its latest efficiency-optimized version~\cite{fang2024mini}, across all seven scenes of the large-scale, real-world NeRF-360 dataset~\cite{barron2022mip}.}
\label{fig:result-runtime}
\end{figure}

\begin{figure}[!t]
\centering
\includegraphics[width=1\linewidth]{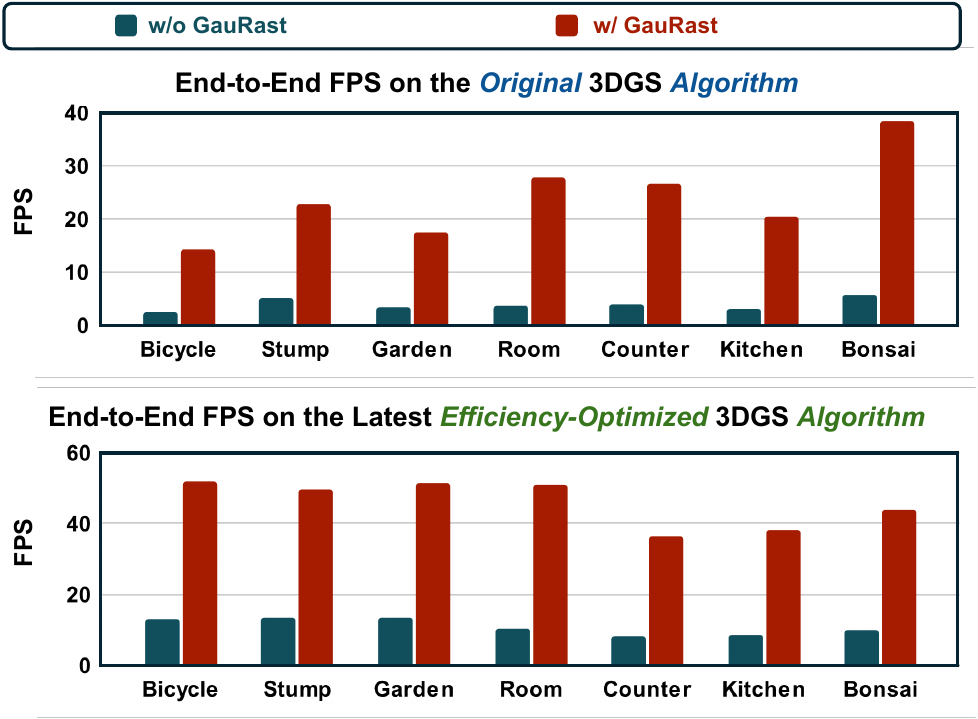}
\vspace{-2em}
\caption{End-to-end FPS comparison between the baseline SoC~\cite{orinnx} and the baseline SoC enhanced with GauRast, using both the original 3DGS~\cite{kerbl20233d} and its latest efficiency-optimized version~\cite{fang2024mini}. The evaluation was conducted across all seven scenes from the large-scale, real-world NeRF-360 dataset~\cite{barron2022mip}.}
\label{fig:result-energy}
\vspace{-1em}
\end{figure}

\subsection{Comparison Against the Baseline SoC}
Fig.~\ref{fig:result-runtime} and \CR{Tab.~\ref{tab:absoluteruntime}} illustrate the speedup and efficiency improvements achieved by GauRast in Gaussian rasterization compared to CUDA implementations across all scenes in the benchmark dataset, for both the original algorithm~\cite{kerbl20233d} and its latest efficiency-optimized version~\cite{fang2024mini}. The results show that our enhanced rasterizer achieves an average runtime reduction of 23$\times$ and an average energy efficiency improvement of 24$\times$ for the original algorithm. For the efficiency-optimized algorithm, it achieves a 20$\times$ runtime reduction and a 22$\times$ improvement in energy efficiency.
Additionally, as shown in Fig.~\ref{fig:result-energy}, replacing CUDA-based rasterization with our proposed enhanced rasterizer leads to substantial end-to-end performance improvements. GauRast achieves a 6$\times$ reduction in runtime for the original 3D Gaussian Splatting algorithm and a 4$\times$ reduction for the optimized algorithm, yielding average frame rates of 24 FPS and 46 FPS, respectively.

\subsection{Comparison Against SOTA 3DGS Accelerator} 
SOTA work GSCore~\cite{lee2024gscore} on 3DGS rendering acceleration achieved a 20$\times$ speedup on an edge SoC~\cite{xnx} for Gaussian rasterization, utilizing a dedicated area of \qty{3.95}{\milli\meter\squared} with FP16 precision. 
If we re-implement GauRast to perform FP16 operations, our design achieves equivalent performance to GSCore but requires only \qty{0.16}{\milli\meter\squared} area, achieving a 24.7$\times$ improvement in area efficiency. This efficiency gain is attributed to the reuse of existing resources built in the triangle rasterizer. These results highlight the effectiveness of our enhancement approach in efficiently leveraging existing GPU components.

\CR{\subsection{Compatibility with non-NVIDIA GPUs}
GauRast is compatible with any GPU equipped with a triangle rasterizer, including but not limited to those from NVIDIA. To further evaluate its generalizability, we conducted an experiment using an Apple M2 Pro GPU~\cite{M2Pro} with OpenSplat~\cite{OpenSplat}. The M2 Pro provides 2.6× greater FP32 computing capability than the baseline NVIDIA Orin NX GPU~\cite{orinnx}. Our results demonstrate that GauRast achieves an 11.2× speedup in rasterization on the bicycle scene, confirming its compatibility across different GPU architectures.}
\section{Conclusion}
This work presents GauRast, an enhancement for existing GPU rasterizers designed to accelerate 3DGS rendering. Experiment results show that our enhanced rasterizer achieves a 6$\times$ speedup in end-to-end runtime for the original 3DGS and a 4$\times$ speedup for the latest efficiency-optimized pipeline on the NVIDIA Jetson Orin NX, with an area overhead of only 0.2\%, demonstrating a promising path forward to enable real-time 3DGS rendering on edge devices.